\documentclass[twocolumn,superscriptaddress,amsmath,amssymb,aps,prb]{revtex4-2}
\usepackage{graphicx}
\usepackage{CJKutf8}
\usepackage{dcolumn}
\usepackage{amsmath,bm}
\usepackage{epstopdf}
\usepackage[mathlines]{lineno}
\usepackage{color}
\usepackage[colorlinks=true,linkcolor=blue,citecolor=magenta,urlcolor=cyan]{hyperref}
\usepackage{orcidlink}

\begin{document}
\title{Low-energy effective Hamiltonian and end states of an inverted HgTe nanowire}
\author{Rui\! Li~(\begin{CJK}{UTF8}{gbsn}李睿\end{CJK})\,\orcidlink{0000-0003-1900-5998}}
\email{ruili@ysu.edu.cn}
\affiliation{Hebei Key Laboratory of Microstructural Material Physics, School of Science, Yanshan University, Qinhuangdao 066004, China}

\begin{abstract}
{\color{red}}
The band inversion transition in a cylindrical HgTe nanowire is inducible via varying the nanowire radius. Here we derive the low-energy effective Hamiltonian describing the band structure of the HgTe nanowire close to the fundamental band gap. Because both the $E_{1}$ and $H_{1}$ subbands have quadratic dependence on $k_{z}$ when the gap closes, we need to consider at least three subbands, i.e., the $E_{1}$, $H_{1}$, and $H_{2}$ subbands, in building the effective Hamiltonian. The resulting effective Hamiltonian is block diagonal and each block is a $3\times3$ matrix. End states are found in the inverted regime when we solve the effective Hamiltonian with open boundary condition.
\end{abstract}
\date{\today}
\maketitle

\section{Introduction}
{\color{red}}
The Thouless--Kohmoto--Nightingale--Nijs index~\cite{PhysRevLett.49.405}, originally developed for explaining the quantization of the Hall conductance~\cite{PhysRevLett.45.494}, also serves to distinguish the Chern insulator from the ordinary insulator~\cite{PhysRevLett.61.2015,bernevig2013topological,shen2012topological}. However, the Thouless--Kohmoto--Nightingale--Nijs index vanishes in time-reversal invariant systems, Kane and Mele discovered that a $Z_{2}$ index is needed to distinguish the time-reversal-invariant topological insulator from the ordinary insulator~\cite{PhysRevLett.95.226801,PhysRevLett.95.146802}. A change of the topological index by tuning some system parameters is usually accompanied with a gap-closing-and-reopening transition in the band structure~\cite{RevModPhys.82.3045,RevModPhys.83.1057}. There exist edge states in one of the gapped phases, i.e., the topologically nontrivial phase, in the presence of open boundary condition~\cite{PhysRevB.74.085308,PhysRevLett.101.246807}. Two-dimensional topological insulator was theoretically predicted in graphene~\cite{PhysRevLett.95.226801} and experimentally demonstrated in HgTe quantum well~\cite{doi:10.1126/science.1133734,doi:10.1126/science.1148047}. The three-dimensional generalizations of the two-dimensional topological insulator were established soon~\cite{PhysRevLett.98.106803,PhysRevB.75.121306,PhysRevB.79.195322}, and several realistic materials were predicted~\cite{PhysRevB.76.045302} and confirmed as three-dimensional topological insulators~\cite{Hsieh:2008tf,Xia:2009uh,Zhang:2009vu,doi:10.1126/science.1173034,PhysRevLett.106.126803}.

Although the polyacetylene described using the Su-Schrieffer-Heeger model~\cite{PhysRevLett.42.1698} can be regarded as a one dimensional topological insulator, the searching for realistic topological insulator in other quasi-one-dimensional materials is still of fundamental importance~\cite{PhysRevLett.83.2636,Guo:2016wi}. Inspired by the successful discovery of the topological insulator in a quasi-two-dimensional HgTe quantum well~\cite{doi:10.1126/science.1133734}, one should also seek the one-dimensional topological insulator in those semiconductor materials with a negative band gap~\cite{madelung2004semiconductors,Hsieh:2012tq,PhysRevB.72.035321,Orlita:2014tw}. Encouraging results were obtained in SnTe~\cite{doi:10.1021/acsanm.4c00506} and HgTe~\cite{RL_2025a} nanowires, where signatures of the band inversion transition were demonstrated via tuning the nanowire size or radius~\cite{doi:10.1021/acsanm.4c00506,RL_2025a}.

When the nanowire is in the inverted regime, i.e., the topologically nontrivial phase, there should exist end states. Here, we derive the low-energy effective Hamiltonian of the HgTe nanowire near the topological transition point $R_{c}=3.2$ nm~\cite{RL_2025a}, and study the potential end states in the inverted regime. We find the minimal effective Hamiltonian in each spin subspace is represented by a $3\times3$ matrix, which differs from the usual two-band Dirac description in the HgTe quantum well~\cite{doi:10.1126/science.1133734}. This difference originates from the fact that, the dispersions are quadratic in $k_{z}$ near the gap-closing-and-reopening transition in the nanowire~\cite{RL_2025a}, while they are linear in $k_{\parallel}$ (Dirac dispersions) in the quantum well~\cite{doi:10.1126/science.1133734}. Solving the lower-energy effective Hamiltonian in the inverted regime with open boundary condition, we indeed find four end states in total.  There are two end states (spin up and spin down) localized at each end of the nanowire. The energies of the end states lie in the subband gap, and are much closer to the conduction band edge.

\section{Effective mass model}
{\color{red}}
\begin{figure}
\includegraphics[width=8.5cm]{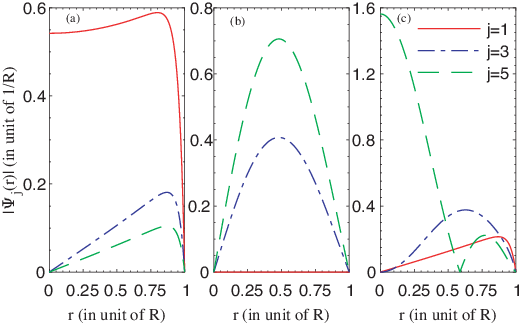}
\caption{\label{fig_eigenstate}The components of the wavefunction at $k_{z}=0$ as a function of the coordinate $r$. The HgTe nanowire has a radius of $R=3.4$ nm. The results of the $|E_{1},1/2\rangle$ state (a), $|H_{1},1/2\rangle$ state (b), and the $|H_{2},-1/2\rangle$ state (c). }
\end{figure}

Although we focus on the low-energy effective Hamiltonian of the HgTe nanowire in this paper, some necessary information and conclusions about the nanowire model studied in our previous paper~\cite{RL_2025a} should be briefly reviewed. A cylindrical HgTe nanowire placed along the $z$ direction is under consideration, the subband quantization of the nanowire is governed by the following effective mass Hamiltonian
\begin{equation}
H=H_{\rm Kane}({\bf k}\rightarrow-i\nabla)+V(r),\label{eq_effmassmodel}
\end{equation}
where $H_{\rm Kane}({\bf k})$ is the six-band Kane model~\cite{KANE1957249} covering both the $\Gamma_{6}$ and $\Gamma_{8}$ bands of HgTe, and $V(r)$ is a hard-wall confining potential
\begin{equation}
V(r)=\left\{\begin{array}{cc}0,&r<R,\\\pm\infty,&r>R,\end{array}\right.\label{eq_confining}
\end{equation}
with $R$ being the radius of the nanowire. The spin-orbit split-off $\Gamma_{7}$ band is 1.08 eV in energy lower than the $\Gamma_{8}$ band, and its contribution to the low-energy subband dispersions in the HgTe quantum well was estimated to be less than 5\%~\cite{Pfeuffer2000,doi:10.1126/science.1133734}. Here for simplicity we also neglect the contribution from the $\Gamma_{7}$ band in the nanowire. Because of the negative band gap, the six-band Kane model can be regarded as the minimal bulk model of HgTe. Note that we have used a cylindrical coordinate representation ${\bf r}=(r,\varphi,z)$ and the nanowire axis is defined as the $z$ axis. The positive sign in Eq.~(\ref{eq_confining}) applies to the $\Gamma_{6}$ band and the negative sign applies to the $\Gamma_{8}$ band. When using the six-band instead of the eight-band Kane model, the removing of the $\Gamma_{7}$ band leads to a renormalization of the mass in the $\Gamma_{6}$ band~\cite{winkler2003spin}. Here we neglect this small renormalization from the $\Gamma_{7}$ band, the main contribution to the mass in the $\Gamma_{6}$ band comes from the $\Gamma_{8}$ band, and is naturally included in the six-band Kane model. The detailed bulk band parameters of HgTe are given in the table of our previous paper~\cite{RL_2025a}.

In order to analytically tackle the problem of subband quantization in the nanowire, we take the spherical approximation in the $\Gamma_{8}$ band of the Kane model~\cite{PhysRev.102.1030}. We replace the Luttinger parameters $\gamma_{2,3}$ with the weighted average $\gamma_{s}= (2\gamma_{2} + 3\gamma_{3})/5$ in order to minimize the effects of the non-spherical terms~\cite{PhysRevB.8.2697}. The relatively small terms proportional to $\gamma_{2,3}-\gamma_{s}$ are neglected in the current study. Note that in calculating the hole subbands and deriving the low-energy effective Hamiltonian of a Ge nanowire, the spherical approximation yields results that are qualitatively consistent with those from more accurate methods~\cite{PhysRevB.84.195314,RL2023c,RL_2025b}. Under the spherical approximation~\cite{PhysRev.102.1030}, the effective mass Hamiltonian (\ref{eq_effmassmodel}) is exactly solvable~\cite{RL_2025a,Li-lirui:2024we} following the method of Sercel and Vahala~\cite{PhysRevB.42.3690}. The conserved quantity $F_{z}$, i.e., the $z$-component of the total angular momentum, can be used to classify the eigenstates. The eigenstates at $k_{z}=0$ are important for constructing the low-energy effective Hamiltonian~\cite{RL_2025b}. The $6\times6$ Hamiltonian (\ref{eq_effmassmodel}) at $k_{z}=0$ can be written as a block diagonal form with two independent $3\times3$ blocks. In one block, the eigenfunction can be written as 
\begin{equation}
\Psi_{F_{z}}(r,\varphi)=\left(\begin{array}{c}\Psi_{1}(r)e^{im\varphi}\\0\\\Psi_{3}(r)e^{i(m-1)\varphi}\\0\\\Psi_{5}(r)e^{i(m+1)\varphi}\\0\end{array}\right),\label{eq_wavefunc1}
\end{equation}
where $F_{z}=m+1/2$ ($m=0,\pm1,\pm2\ldots$). In the other block, we have
\begin{equation}
\Psi_{F_{z}}(r,\varphi)=\left(\begin{array}{c}0\\\Psi_{2}(r)e^{i(m+1)\varphi}\\0\\\Psi_{4}(r)e^{im\varphi}\\0\\\Psi_{6}(r)e^{i(m+2)\varphi}\end{array}\right).\label{eq_wavefunc2}
\end{equation}
Therefore, the eigenfunctions at $k_{z}=0$ in general have three vanishing components, i.e., either $\Psi_{2,4,6}(r)=0$ or $\Psi_{1,3,5}(r)=0$.

Via tuning the radius $R$ of the nanowire, the lowest electron subband $E_{1}$ intersects with the two top most hole subbands $H_{1}$ and $H_{2}$~\cite{RL_2025a}. The band inversion transition is thus inducible by the nanowire radius. Here we show the wavefunctions of the eigenstates $|E_{1},1/2\rangle$, $|H_{1},1/2\rangle$, and $|H_{2},-1/2\rangle$ in Figs.~\ref{fig_eigenstate}(a), (b), and (c), respectively. These eigenstates have the form of Eq.~(\ref{eq_wavefunc1}), while the eigenstates $|E_{1},-1/2\rangle$, $|H_{1},-1/2\rangle$, and $|H_{2},1/2\rangle$ have the form of Eq.~(\ref{eq_wavefunc2}) (not shown). The $|E_{1},1/2\rangle$ state is an interface state~\cite{PhysRevB.31.2557,PhysRevB.32.5561,doi:10.1126/science.1133734}, i.e., the interface between the HgTe material and the trivial vacuum. The $|H_{1},1/2\rangle$ state does not possess any electron characteristics, i.e., the electron components are zero $\Psi_{1,2}(r)=0$. The $|H_{2},-1/2\rangle$ state is a normal eigenstate with both electron and hole components.

\section{Low-energy effective Hamiltonian}

\begin{table*}
\caption{\label{tab1}Parameters of the effective Hamiltonian (\ref{eq_effHamiltonian}).}
\begin{ruledtabular}
\begin{tabular}{cccccccccc}
$R$ (nm)&$A$~(eV)&$B$~(eV)&$C$~(eV)&$D$~(eV)&$F$~(eV)&$G$~(eV)&$\Delta_{e}$~(eV)&$\Delta_{h1}$~(eV)&$\Delta_{h2}$~(eV)\\
3.0&0.0029&-0.0215&-0.0188&-0.0016$i$&0.0855&-0.0526$i$&-0.1033&-0.1330&-0.2094\\
3.2&0.0025&-0.0187&-0.0166&-0.0014$i$&0.0647&-0.0459$i$&-0.1156&-0.1156&-0.1875\\
3.4&0.0023&-0.0167&-0.0150&-0.0013$i$&0.0507&-0.0408$i$&-0.1245&-0.1036&-0.1714
\end{tabular}
\end{ruledtabular}
\end{table*}

\begin{figure*}
\includegraphics{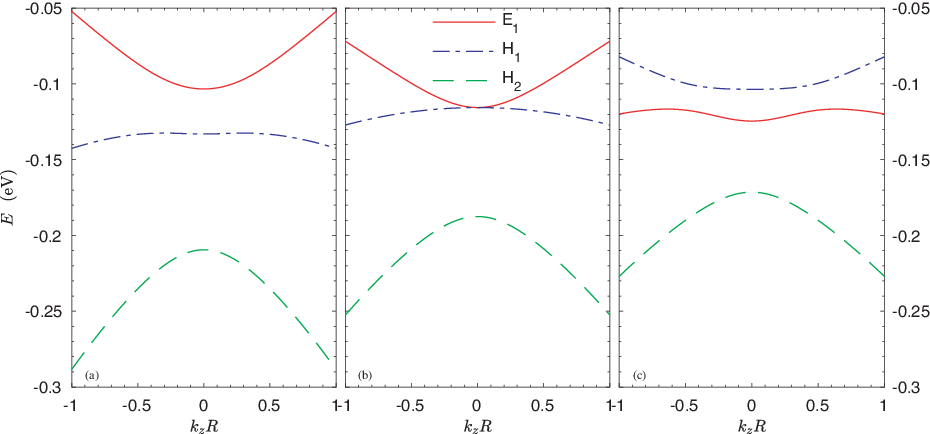}
\caption{\label{fig_effband}The $E_{1}$, $H_{1}$, and $H_{2}$ subband dispersions near $k_{z}=0$. (a) Result for radius $R=3.0$ nm with normal band structure. (b) Result for radius $R=3.2$ nm with critical band structure. (c) Result for radius $R=3.4$ nm with inverted band structure.}
\end{figure*}

{\color{red}}
The low-energy effective Hamiltonian describing the subband dispersions close to the fundamental band gap can be built using the standard method~\cite{winkler2003spin}. The $k_{z}$ terms in the effective mass Hamiltonian (\ref{eq_effmassmodel}) are treated as perturbation~\cite{PhysRevB.84.195314,RL_2025b}. We first solve both the eigenvalues and eigenstates of $H(k_{z}=0)$, and the eigenstates (see Fig.~\ref{fig_eigenstate}) serve as the basis states for building the effective Hamiltonian. We then calculate the matrix elements of $H(k_{z})$ by choosing a proper set of basis states, and eventually $H(k_{z})$ can be written as an effective Hamiltonian. Because $k_{z}$ is treated perturbatively, the effective Hamiltonian is expected to be valid only for small $k_{z}$'s, i.e., $k_{z}R\ll1$.

The band inversion transition occurred in the nanowire is very similar to that in the quantum well, e.g., the $E_{1}$ state is an interface state, the $H_{1}$ state has only hole characteristics, and the critical sizes are almost the same in both cases, i.e., $2R_{c}=6.4$ nm~\cite{doi:10.1126/science.1133734,RL_2025a}.  However, at the gap-closing-and-reopening transition point, i.e., at the critical radius $R_{c}\approx3.2$ nm, both the $E_{1}$ and $H_{1}$ subbands in the HgTe nanowire have quadratic dependence on $k_{z}$~\cite{RL_2025a}. This result differs apparently from that in a quasi-two-dimensional HgTe quantum well where both the $E_{1}$ and $H_{1}$ subbands have linear dependence on $k_{\parallel}$~\cite{doi:10.1126/science.1133734}.  As a consequence, in building the low-energy effective Hamiltonian we need to involve at least three subbands, i.e., the $E_{1}$, $H_{1}$, and $H_{2}$ subbands. Note that due to the conservation of the $z$-component of the total angular momentum $F_{z}$~\cite{RL_2025a}, subbands labeled with different values of $F_{z}$ are naturally decoupled in building the effective Hamiltonian. Our previous paper also showed that for radius $2.5<R<4.0$ {\rm nm}, the subbands $E_{1}$, $H_{1}$, and $H_{2}$ are not only close to the fundamental band gap but are well separated from other subbands of $|F_{z}|=1/2$~\cite{RL_2025a}. In the low-energy Hilbert subspace spanned by the ordered set $|E_{1},1/2\rangle$, $|H_{1},1/2\rangle$, $|H_{2},1/2\rangle$, $|E_{1},-1/2\rangle$, $|H_{1},-1/2\rangle$, and $|H_{2},-1/2\rangle$, we obtain the following $6\times6$ effective Hamiltonian 
\begin{equation}
H_{\rm ef}(k_{z})=\left(\begin{array}{cc}H_{+}(k_{z})&0\\0&H^{*}_{+}(-k_{z})\end{array}\right),\label{eq_effHamiltonian}
\end{equation}
where
\begin{equation}
H_{+}(k_{z})=\left(\begin{array}{ccc}Ak^{2}_{z}+\Delta_{e}&Dk^{2}_{z}&Fk_{z}\\
D^{*}k^{2}_{z}&Bk^{2}_{z}+\Delta_{h1}&Gk_{z}\\
F^{*}k_{z}&G^{*}k_{z}&Ck^{2}_{z}+\Delta_{h2}\end{array}\right),\label{eq_efHamil}
\end{equation}
with $k_{z}$ in units of $1/R$. The parameters $A$, $B$, $C$, $D$, $F$, $G$, $\Delta_{e}$, $\Delta_{h1}$, and $\Delta_{h2}$ depend on the nanowire radius $R$. The values of these parameters near the critical radius $R_{c}=3.2$ nm are given in Tab.~\ref{tab1}. The form of $H^{*}_{+}(-k_{z})$ in the lower block of $H_{\rm ef}$ is a consequence of the time-reversal symmetry~\cite{doi:10.1126/science.1133734}. We can see in the $E_{1}$ and $H_{1}$ subspace, the dispersions are indeed quadratic in $k_{z}$ (see the first $2\times2$ block matrix of Eq.~(\ref{eq_efHamil})).

The low-energy effective Hamiltonian (\ref{eq_effHamiltonian}) indeed qualitatively captures the dispersion behaviors near $k_{z}=0$~\cite{RL_2025a}. We show the subband dispersions calculated using the effective Hamiltonian in Figs.~\ref{fig_effband}(a), (b), and (c) for nanowire radii R=3.0 nm, 3.2 nm, and 3.4 nm, respectively. For radius $R=3.0$ nm, the nanowire is in the normal regime with the $E_{1}$ subband lying above the $H_{1}$ subband, i.e., $\Delta_{e}>\Delta_{h1}$. For radius $R=3.2$ nm, the nanowire is in the critical regime with the $E_{1}$ subband touching with the $H_{1}$ subband, i.e., $\Delta_{e}=\Delta_{h1}$. For radius $R=3.4$ nm, the nanowire is in the inverted regime with the $E_{1}$ subband lying below the $H_{1}$ subband, i.e., $\Delta_{e}<\Delta_{h1}$.

\section{End states}

\begin{figure}
\includegraphics{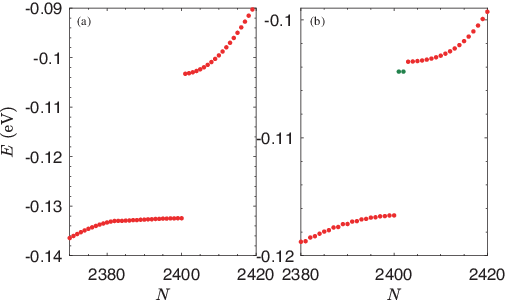}
\caption{\label{fig_openspectrum}The energy as a function of the state number in a HgTe nanowire with finite length 400 nm. (a) Result for radius $R=3.0$ nm. (b) Result for radius $R=3.4$ nm. There is no energy state in the subband gap for radius $R=3.0$ nm, while there exist two energy states (marked green) in the subband gap for radius $R=3.4$ nm. }
\end{figure}

\begin{figure}
\includegraphics{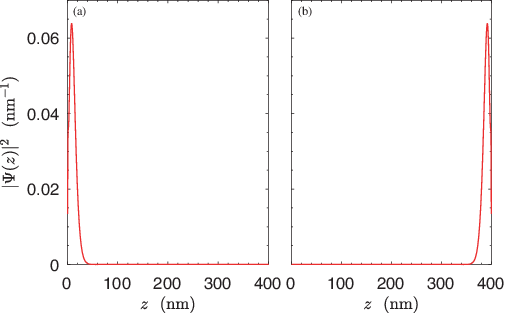}
\caption{\label{fig_edgestate}The probability density distribution of the two energy states in the subband gap shown in Fig.~\ref{fig_openspectrum}(b). The left end (a) and the right end (b) states in the inverted HgTe nanowire. The distance of the end states extending into the bulk is about 40 nm.}
\end{figure}

{\color{red}}
The gap-closing-and-reopening transition in the band structure is a signature of the topological phase transition. Previous studies have demonstrated similar band-inversion transition in various physical systems~\cite{Kitaev:2001up,PhysRevLett.83.2636,doi:10.1126/science.1133734,PhysRevB.74.085308,Guo:2016wi}. However, most reported transitions are characterized by the Dirac Hamiltonian~\cite{shen2012topological}. It is believed that one of the gapped phases must be topologically nontrivial. We guess the HgTe nanowire in the inverted regime is actually a topological insulator. A hallmark of the topological insulator is the existence of edge or end states~\cite{PhysRevLett.61.2015,PhysRevLett.95.226801}. Here we consider a HgTe nanowire with finite length in order to study its potential end states. The Hamiltonian under investigation reads
\begin{equation}
H_{\rm open}=H_{+}(k_{z}\rightarrow-i\partial_{z})+V(z),
\end{equation}
where $V(z)=0$ for $0<z<400$ nm, and $V(z)=\pm\infty$ for elsewhere, i.e., the nanowire has a finite length of 400 nm. We insert 1200 equidistant sites into the interval $0<z_{1}<z_{2}\ldots\,z_{1200}<400$ nm, and solve the eigenvalues of $H_{\rm open}$ using finite difference method. 

In Fig.~\ref{fig_openspectrum}, we plot the energies in the nanowire as a function of the state number. Note that we only focus on those energy states near the fundamental subband gap. In the normal regime $R=3.0$ nm, we do not find energy states in the subband gap (see Fig.~\ref{fig_openspectrum}(a)). While in the inverted regime $R=3.4$ nm, we indeed find two energy states in the subband gap (see the dots marked green in Fig.~\ref{fig_openspectrum}(b)). The energies of these two energy states are much closer to the conduction band edge. 

In order to confirm the two energy states in the subband gap are actually end states, we calculate the probability density distribution of these two states. One state is localized at the left end of the nanowire (see Fig.~\ref{fig_edgestate}(a)), while the other state is localized at the right end of the nanowire (see Fig.~\ref{fig_edgestate}(b)). Thus, the inverted HgTe nanowire is indeed a one-dimensional topological insulator. When the lower block $H^{*}_{+}(-k_{z})$ of the effective Hamiltonian is considered similarly, we totally have four end states. There are two states localized at each end of the nanowire, and they are related to each other via time-reversal symmetry. 

If one wants to use the $2\times2$ effective Hamiltonian through dimension reduction~\cite{winkler2003spin}. For example, one can use quasi-degenerate perturbation theory to eliminate the $H_{2}$ subband and obtain 
\begin{equation}
H_{+}(k_{z})=\left(\begin{array}{cc}A'k^{2}_{z}+\Delta_{e}&D'k^{2}_{z}\\
D'^{*}k^{2}_{z}&B'k^{2}_{z}+\Delta_{h1}\end{array}\right),\label{eq_efHamil2}
\end{equation}
where 
\begin{eqnarray}
A'&=&A+\frac{|F|^{2}}{\Delta_{e}-\Delta_{h2}},\nonumber\\
B'&=&B+\frac{|G|^{2}}{\Delta_{h1}-\Delta_{h2}},\nonumber\\
D'&=&D+\frac{FG^{*}}{2}\left(\frac{1}{\Delta_{e}-\Delta_{h2}}+\frac{1}{\Delta_{h1}-\Delta_{h2}}\right).
\end{eqnarray}
However, repeating the finite difference calculations using the above $2\times2$ Hamiltonian, we do not find end states in both of the gapped phases. The minimal effective Hamiltonian in one spin subspace is thus a $3\times3$ matrix introduced by Eq.~(\ref{eq_efHamil}). The situation here is similar to that in the bulk $\Gamma_{8}$ valence bands, where the couplings between the heavy hole and the light hole bands are quadratic in $k$~\cite{winkler2003spin}. Here the coupling between $E_{1}$ and $H_{1}$ subbands is also quadratic in $k_{z}$. We need to consider an additional subband (here is the $H_{2}$ subband) that has a linear $k_{z}$ coupling to both the $E_{1}$ and $H_{1}$ subbands, similar to the case of the bulk ${\bf k}\cdot{\bf p}$ model involving both the $\Gamma_{6}$ and $\Gamma_{8}$ bands. In this sense, the topological phase demonstrated in our paper is somewhat  similar to that in strained bulk HgTe or $\alpha$-Sn~\cite{PhysRevB.76.045302,PhysRevB.77.125319,PhysRevLett.106.126803}.

\section{Discussion and summary}
Due to the large surface-to-volume ratio, the lattice mismatched strain can be elastically relaxed in epitaxially grown nanowire. Therefore, we do not consider the strain effects in this paper. We note that strain has more profound influence on the subband structure in the nanowire than in the quantum well, e.g., the strain can significantly change the shape of the hole subband dispersions in a Ge nanowire~\cite{PhysRevB.84.195314}. 

In summary, we have derived the low-energy effective Hamiltonian describing the band structure close to the fundamental band gap for a cylindrical HgTe nanowire. In each of the spin subspaces, the minimal effective Hamiltonian has a dimension of three and differs from the usual one-dimensional Dirac Hamiltonian. Via solving the effective Hamiltonian in the presence of the open boundary condition, end states are found in the inverted regime. If the effective Hamiltonian is reduced from the $3\times3$ form to the $2\times2$ form using the dimension reduction method, the end state physics are also lost in this reduction. 

\bibliography{Ref_RL}
\end{document}